\documentclass[showpacs,nofootinbib,preprintnumbers,amsmath,amssymb,preprint,aps]{revtex4}
\usepackage[dvips]{graphicx}
\usepackage{graphicx}
\usepackage{amsfonts}
\usepackage{bm}
\usepackage{amsmath}
\usepackage{amssymb}
\usepackage{color}
\usepackage[all]{xy}

\def\be{\begin{equation}}
\def\ee{\end{equation}}
\def\bea{\begin{eqnarray}}
\def\eea{\end{eqnarray}}

\begin{document}

\title{Homogeneity and thermodynamic identities in geometrothermodynamics}

\author{Hernando  Quevedo$^{1,2}$, Mar\'\i a N. Quevedo$^3$ and Alberto S\'anchez$^4$}
\email{quevedo@nucleares.unam.mx,maria.quevedo@unimilitar.edu.co,asanchez@nucleares.unam.mx}
\affiliation{
$^1$Instituto de Ciencias Nucleares, Universidad Nacional Aut\'onoma de M\'exico,
 AP 70543, M\'exico, DF 04510, Mexico\\
$^2$Dipartimento di Fisica and ICRANet, Universit\`a di Roma ``La Sapienza",  I-00185 Roma, Italy\\
$^3$ Departamento de Matem\'aticas, 
 Facultad de Ciencias B\'asicas, 
Universidad Militar Nueva Granada, Cra 11 No. 101-80, 
Bogot\'a D.E., Colombia \\
$^4$Departamento de Posgrado, CIIDET,  AP752, 
Quer\'etaro, QRO 76000, Mexico}

\date{\today}

\begin{abstract}

We propose a classification of thermodynamic systems in terms of the homogeneity properties of their fundamental equations. 
Ordinary systems correspond to homogeneous functions and non-ordinary systems are given by generalized homogeneous functions.
This affects the explicit form of the Gibbs-Duhem relation and Euler's identity. We show that these generalized relations can be implemented in the formalism of black hole geometrothermodynamics in order to completely fix the arbitrariness present in Legendre invariant metrics. 


\end{abstract}

%\pacs{05.70.Ce; 05.70.Fh; 04.70.-s; 04.20.-q}

\maketitle

\section{Introduction}
\label{sec:int}

In the search for a geometric representation of thermodynamics \cite{amari85} several methods have been proposed. 
The first step to construct a geometric background
consists in associating a particular space to a given thermodynamic system. The obvious candidate is the equilibrium space in which 
each point represents a state of equilibrium of the system. At any moment, the system  occupies a particular point of the equilibrium space. If the system undergoes a quasi-static process, in the equilibrium space it corresponds to a particular path in which each point
represents an equilibrium state. The second step consists in endowing the equilibrium space with a metric. Although there are several possibilities to do that, only Riemannian metrics have been extensively considered, giving rise to the area which is now known as thermodynamic geometry. 

Riemannian geometry was first introduced in statistical physics and thermodynamics
by Rao \cite{rao45}, who proposed a metric the components of which  coincide in local coordinates with Fisher's
information matrix. Rao's original work has been followed up and extended by a number of
authors and the metric is now known as the Fisher-Rao metric (see, for instance, \cite{amari85} for a review).
On the other hand, since any thermodynamic system is completely specified by means of its fundamental equation, either in the entropy or energy representation \cite{callen}, in thermodynamic geometry the metric is usually taken as the Hessian of the fundamental equation. 
In this context, Weinhold \cite{weibook} and Ruppeiner \cite{ruprev} used the Hessian of the internal energy and the entropy, respectively:
\be
g^W = \frac{\partial ^2  U }{\partial E^a \partial E^b} d E^a dE^b  \ , \quad 
g^R = -\frac{\partial ^2  S }{\partial E^a \partial E^b} d E^a dE^b  \ , 
\ee
where $E^a$ ($a=1,...,n)$ represent the extensive variables from which the thermodynamic potentials $U$ and $S$ depend. It can be shown that the Ruppeiner metric is conformally related to the Weinhold metric with the inverse of the temperature as the conformal factor. Moreover, one can introduce in the equilibrium space another metric as the Hessian of any thermodynamic potential that can be obtained 
from $U$ or $S$ by means of a Legendre transformation \cite{chinosmet}.
In general, we can say that the formalism of thermodynamic geometry utilizes the fact that to any system corresponds a thermodynamic potential 
from  which a Hessian can be obtained that is then used to introduce a Riemannian metric into the equilibrium manifold.

The formalism of geometrothermodynamics (GTD) is different because it does not assume any particular Hessian metric for the equilibrium space \cite{quev07}. Instead, the aim of GTD is to incorporate in a geometric way the well-known fact that in classical thermodynamics the physical properties of a system do not depend on the choice of thermodynamic potential \cite{callen}. Since different 
thermodynamic potentials are related by Legendre transformations, in GTD we consider the invariance under Legendre transformations as the main condition to be satisfied by all the geometric structures entering the formalism. It is then expected that the metric for the equilibrium space emerges as a result of imposing Legendre invariance. To this end, it is necessary to represent Legendre transformations as coordinate transformations in the following way \cite{arnold}: Let $Z^A=\{\Phi, E^a, I^a \}$ be the coordinates of a $(2n+1)-$dimensional space ${\cal T}$, where a particular coordinate transformation $Z^A\rightarrow \tilde Z ^A = \{\tilde \Phi, \tilde E^a, \tilde I^a \}$ 
corresponds to a Legendre transformation. The important point about the space ${\cal T}$ is that, according to Darboux theorem, there exists a canonical 1-form $\Theta = d\Phi - I_a d E^a$, $I_a = \delta_{ab} I^b$, such that $\Theta\wedge (d\Theta)^n \neq 0$, i.e., $\Theta$ is a contact form on ${\cal T}$. Also, one can prove that $\Theta$ is Legendre invariant in the sense that under a Legendre transformation it behaves as $\Theta\rightarrow \tilde \Theta = d\tilde \Phi - \tilde I_a d\tilde E^a$. Another important feature of this construction is 
that the equilibrium space ${\cal E}$ emerges also in a natural way as a subspace of ${\cal T}$. Indeed, consider the smooth embedding map 
$\varphi: {\cal E}\rightarrow {\cal T}$ such that the pullback  $\varphi^*(\Theta) = 0$, i.e., $d\Phi = I_a dE^a$ on ${\cal E}$, implying 
that $\Phi=\Phi(E^a)$ and $I_a = \frac{\partial \Phi}{\partial E^a}$. We can now interpret $\Phi(E^a)$ as the fundamental equation in terms of the extensive variables $E^a$. Then, the pullback condition $\varphi^*(\Theta)=0$ coincides with the first law of thermodynamics and ${\cal E}$ and ${\cal T}$ can be interpreted as the equilibrium and phase space, respectively. 
Moreover, if we suppose that $G$ is a Riemannian metric on ${\cal T}$, the pullback induces in a natural way a metric $g$ on ${\cal E}$ by means of $g=\varphi^*(G)$. We thus see that in GTD, we only need to specify  the metric $G$ and the fundamental equation $\Phi(E^a)$ 
in order to find all the geometric properties of the equilibrium space ${\cal E}$. 

According to GTD, the phase space is a $(2n+1)-$dimensional Riemannian contact manifold (${\cal T}, \Theta, G)$ with coordinates 
$Z^A=\{\Phi, E^a, I^a \}$. As we have seen, the contact 1-form $\Theta$ is Legendre invariant. For the phase space to have the same property, we must demand that $G$ be invariant under Legendre transformations too. In addition, we impose that in the case of an ideal gas  the metric $g$ is flat in order to interpret the curvature of ${\cal E}$ as a measure of thermodynamic interaction. 
Under these conditions, the most general metrics that are used in GTD can be split into two classes \cite{qq11,qqs16}
\be
 G^{^{I/II}} = (d\Phi - I_a d E^a)^2 + (\xi_{ab} E^a I^b) (\chi_{cd} dE^c dI^d) \ ,
\label{gupap}
\ee
which are invariant under total Legendre transformations and a third class (summation over all repeated indices)
\be	
	\label{GIIIL}
	G^{^{III}}  =(d\Phi - I_a d E^a)^2  +  \left(E_a I_a \right)^{2k+1}  d E^a   d I^a \ , \quad k\in \mathbb{Z}\ ,
\ee
which is invariant with respect to partial Legendre transformations. 
Here $\xi_{ab}$ and $\chi_{ab}$ are diagonal constant $(n\times n)$-matrices. 
If we choose $\chi_{ab} = \delta_{ab}= {\rm diag}(1,\cdots,1)$, the resulting metric $G^{^I}$ can be used to investigate systems with at least one first-order phase transition. Alternatively, for 
$\chi_{ab} = \eta_{ab}= {\rm diag}(-1,\cdots,1)$, we obtain a metric $G^{^{II}}$ which has been used to describe  systems with second-order phase transitions. 

Thus, we see that the only arbitrariness that remains in the GTD metrics is contained in the diagonal matrix $\xi_{ab}$ which has $n$ arbitrary constants. In this work, we will show that in fact this arbitrariness can be fixed, if we take into account the extensivity property of thermodynamic systems. In Sec. \ref{sec:hom}, we propose to classify thermodynamic systems in ordinary and non-ordinary systems.  This classification can be defined exactly in terms of the homogeneity properties of the fundamental thermodynamic equations.  
Then, in Sec. \ref{sec:eul}, we derive the main thermodynamic identities which are used to fix the free parameters of the GTD metrics.
Finally, in Sec. \ref{sec:con}, we explain how to implement our results to avoid non-physical results that can arise when applying 
the GTD formalism to non-ordinary systems.

%%%%%%%%%%%%%%%%%%%%%%%%%%%%%%%%%%%%%%%%%%%%%%%%%%%%%%%%%5

\section{Extensivity of thermodynamic systems}
\label{sec:hom}

In classical thermodynamics, any system can be given in terms of its fundamental equation \cite{callen} which, using the notation introduced
in the previous section corresponds to the function $\Phi(E^a)$, determined by the smooth map $\varphi: {\cal E}\rightarrow {\cal T}$.
Strictly speaking, a function $\Phi(E^a)$ can be a fundamental equation only if all the variables $E^a$ and the thermodynamic potential 
$\Phi$ are extensive, implying that the 
thermodynamic potential $\Phi$ must be identified either with the entropy $S$ or with the internal energy $U$. This means that the fundamental equation can be given only in two different forms which are usually called entropy and energy representations. These two representations are somehow privileged in the sense that they are the only ones involving just extensive variables. For this reason, we 
denote entropy and  energy as fundamental thermodynamic potentials. Legendre transformations allow us to generate from $S$ and $U$ new thermodynamic potentials which we will call Legendre potentials; their main characteristic is that they depend at least on one intensive variable. 

Extensivity is therefore an important physical property of fundamental potentials in classical equilibrium thermodynamics. This property can be expressed as a mathematical condition of the fundamental equation, namely, it must be a homogeneous function of all variables, i.e.,
a rescaling of all the extensive variables is equivalent to a rescaling of the fundamental potential:
\be
\Phi(\lambda E^a) = \lambda^\beta \Phi(E^a) \ ,
\ee
where $\lambda$ is a real constant and $\beta>0$ is the degree of homogeneity. Ordinary thermodynamic systems are usually characterized by $\beta =1$. This means that the total value of an extensive variable in a systems is equal to the sum of all values in each component of the same system. This implies that the fundamental potentials $S$ and $U$ are proportional, for instance, to the number of components and to the volume of the entire system. 
Intensive variables are different because they have the same value in the entire system and in all its components. This means that they are not affected  by a rescaling and, therefore, their degree of homogeneity is zero. 

Non-ordinary thermodynamic systems can be understood in terms of their degree of homogeneity. If $0<\beta<1$, the system is subextensive whereas for $\beta>1$, it is called supraextensive. It can therefore be expected that, in general, intensive variables are characterized by values $\beta\leq 0$. A particularly interesting example of non-ordinary systems are black holes. Indeed, the entropy of a black hole is not 
proportional to the volume, but to the area of the horizon, as postulated by the Bekenstein-Hawking \cite{bek73,haw74,haw75} entropy relation
\be
S = \frac{1}{4} A\ .
\ee
The most general black hole in Einstein-Maxwell theory depends on only three parameters, namely, mass $M$, angular momentum $J$ and electric charge $Q$. Then, the fundamental equation becomes \cite{dav77}
\be
S=\pi \left(2M^2-Q^2 + 2 \sqrt{M^4-J^2-M^2Q^2}\right) \ ,
\label{fekn}
\ee
which is, however, not a homogeneous function, although from a physical point of view all the variables are extensive. Therefore, for non-ordinary thermodynamic systems, we propose to use a generalized definition of extensivity which is represented by generalized homogeneous functions, i.e, functions $\Phi(E^a)$ satisfying the rescaling condition \cite{sta71}
\be 
\Phi(\lambda^{\beta_1} E^1,\ldots, \lambda^{\beta_n}E^n) = \lambda^{\beta_{\Phi}} \Phi(E^1,\ldots, E^n)\ ,
\label{genhom}
\ee
where $\beta_a = (\beta_1, ..., \beta_n)$ are real constants, and $\beta_{\Phi}$ is the degree of generalized homogeneity. It is then easy to see that the fundamental equation (\ref{fekn}) is a generalized homogeneous function of degree $\beta_S$, if the conditions
\be
\beta_J = 2\beta_M\ ,\quad \beta_Q = \beta_M\ , \quad \beta_S = 2 \beta_M \ ,
\label{cond1}
\ee
are satisfied. 

The consistency of this definition of generalized extensivity can be shown by considering the second fundamental potential $M$ 
which can be obtained by inverting the fundamental equation (\ref{fekn}):
\be
M= \left[ \frac{\pi J^2}{S} +  \frac{S}{4\pi}\left(1+\frac{\pi Q ^2}{S}\right)^2 \right]^{1/2} \ .
\ee
This is a generalized homogeneous function of degree $\beta_M$ for the choice
\be
\beta_J = \beta_S\ ,\quad \beta_Q = \frac{1}{2} \beta_S \ ,\quad \beta_M = \frac{1}{2} \beta_S
\ ,
\ee
which is consistent with the conditions (\ref{cond1}).   
Notice that, in general,  extensivity is a property of the fundamental potentials only. Indeed, all the Legendre potentials depend on at least an intensive variable which in the case of ordinary systems is of degree zero. Then, one can rescale only the extensive variables, keeping unchanged all the intensive variables.

%%%%%%%%%%%%%%%%%%%%%%%%%%%%%%%%%%%%%%%%%%%%%%%%%%%%%%%

\section{Thermodynamic identities}
\label{sec:eul}

The extensivity properties of thermodynamic systems lead to a number of identities which are useful for the investigation of 
physical properties. Consider the generalized homogeneous function (\ref{genhom}) and compute the derivative with respect to the
parameter $\lambda$ on both sides of the equation. Evaluating the resulting expression for $\lambda=1$, we obtain
\be
\beta_1 \frac{\partial \Phi}{\partial E^1} E^1 + \cdots + \beta_n \frac{\partial \Phi}{\partial E^n} E^n = \beta_\Phi \Phi \ .
\ee
From the first law $d\Phi = I_a dE^a$, we see that $I_a=\frac{\partial \Phi} {\partial E^a}$. Then, the above equation can be 
expressed as
\be
\beta_{ab} I^a E^b = \beta_\Phi \Phi\ , \quad {\rm  with}\quad \beta_{ab}={\rm diag}(\beta_1,\cdots,\beta_n) \ ,
\label{geneul}
\ee
which in the case of ordinary systems reduces to Euler's identity with $\beta_{ab} = \delta_{ab}$.

In the particular case of black holes, the first law implies that
\be
dM = T dS + \Omega d J + \phi dQ\ ,
\ee
where $T$ is the temperature, $\Omega$ is the angular velocity on the horizon, and $\phi$ is the electric potential. Then, from Euler's 
identity we obtain 
\be
M=2TS + 2 \Omega J + \phi Q \ ,
\ee
which can be recognized as the Smarr formula for black holes \cite{dav77}. 

Computing the derivative of Euler's identity, and using the first law of thermodynamics, we obtain
\be
(\beta_{ab} - \beta_\Phi \delta_{ab}) I^a dE^b + \beta_{ab} E^b dI ^a =0 \ ,
\ee
which is the Gibbs-Duhem relation for non-ordinary systems. In the case of ordinary systems with degree of homogeneity $\beta_\Phi=1$ 
and $\beta_{ab}=\delta_{ab}$, we obtain the standard result. 

We now investigate how the thermodynamic identities can  be used in the GTD metrics. As mentioned above, the metric $g$ of the equilibrium space ${\cal E}$ is induced canonically by means of the pullback $\varphi^*(G)=g$. Then, from the metric (\ref{gupap}), we obtain
\be
g^{^{I/II}} = (\xi_{ab} \delta ^{bc} E^a \Phi_{,c})( \chi_a^{\ b} \Phi_{,bc} dE^adE^c) \ ,
\ee
with 
\be 
\chi_a^{\ b} = \chi_{ac} \delta^{cb}\ , \quad \Phi_{,a} = \frac{\partial \Phi}{\partial E^a} \ .
\ee
The arbitrariness is contained in the conformal factor of the metric $g^{^{I/II}}$. In the case of ordinary systems, we can choose 
$\xi_{ab}=\delta_{ab}$ so that the conformal factor becomes $\beta \Phi$. This is the choice that has been used to analyze several examples
in the GTD formalism. In the case of non-ordinary systems, we can still use $\xi_{ab}=\delta_{ab}$, in which case the conformal factor is not
necessarily proportional to the fundamental potential $\Phi$. Therefore, to obtain a general result which is valid in all possible cases, we choose
\be
\xi_{ab}={\rm diag}(\beta_1,\cdots,\beta_n) \ ,
\ee
which together with the generalized Euler identity (\ref{geneul}) implies that 
 the general metric for the equilibrium space becomes
\be
g^{^{I/II}} = \beta_\Phi \Phi\, ( \chi_a^{\ b} \Phi_{,bc} dE^a dE^c) \ ,
\ee
which is a metric with no arbitrary constants at all. 

We thus see that the $n$ constants contained in $\xi_{ab}$, which the Legendre invariance condition leaves arbitrary, become now fixed 
and correspond to the degree of homogeneity of the extensive thermodynamic variables.

%%%%%%%%%%%%%%%%%%%%%%%%%%%%%%%%%%%%%%%%%%%%%%%%%%%%%%%%

\section{Conclusions}
\label{sec:con}
 
In this work, we use the extensivity property of thermodynamic systems in order to classify them into ordinary and non-ordinary systems. 
The first class are represented by fundamental equations which correspond to homogeneous functions, whereas systems of the second class are 
described by generalized homogeneous functions. This classification leads to a set a generalized thermodynamic identities which relate the thermodynamic potentials and its derivatives. We use in particular the generalized Euler identity to fix the only remaining arbitrariness of the Legendre invariant metrics used in GTD. As a result, we obtain that all the GTD metrics that are invariant under total Legendre transformations induce in the equilibrium space conformal metrics which contain essentially the thermodynamic potential in 
the conformal factor.

In a previous work \cite{azr14}, it was pointed out that some black hole configurations, when analyzed within the framework of the GTD formalism, are characterized by a phase transition structure which does not coincide with the one obtained in black hole thermodynamics. 
The results obtained in the present work clarify all the particular cases analyzed in \cite{azr14}. Indeed, since all the components of the GTD metric used to describe black hole configurations are in general proportional to the thermodynamic potential $\Phi$, any curvature singularity which follows when the condition $\Phi=0$ is satisfied can be considered as unphysical because it corresponds to a configuration with no mass or no entropy. 

Another point criticized in \cite{azr14} was regarding the non-homogeneity of a particular Legendre potential. As we have shown here, in general a Legendre potential cannot be  given in terms of a homogeneous functions, because it contains at least one intensive variable which cannot be rescaled as an extensive variable.

\section*{Acknowledgements}

This work was carried out within the scope of the project CIAS 2045
supported by the Vicerrector\'\i a de Investigaciones de la Universidad
Militar Nueva Granada - Vigencia 2016. This work was partially supported
by UNAM-DGAPA-PAPIIT, Grant No. 111617.

%******************************** ***********************************************************************

\end{document}